\documentclass{llncs}

\usepackage{graphicx}
\usepackage{tabularx}
\usepackage{hyperref}
\usepackage{graphicx}
\usepackage[british]{babel}
\usepackage{eurosym}
\usepackage{pbox}
\usepackage{url}
\usepackage{xspace}
\makeatletter
\g@addto@macro{\UrlBreaks}{\UrlOrds}
\usepackage{booktabs}
\usepackage[misc]{ifsym}
\usepackage[font=small,labelfont=bf,labelsep=space]{caption}
\newcommand\blfootnote[1]{%
	\begingroup
	\renewcommand\thefootnote{}\footnote{#1}%
	\addtocounter{footnote}{-1}%
	\endgroup
}

\usepackage{geometry} 

\begin{document}
\mainmatter

\title{Cybercrime and You: How Criminals Attack and the Human Factors That They Seek to Exploit}
\titlerunning{Cybercrime and You}
 
\author{Jason R. C. Nurse}
\authorrunning{Nurse}

\institute{School of Computing, University of Kent, UK\\
\email{j.r.c.nurse@kent.ac.uk}}

\maketitle

\begin{abstract}
Cybercrime is a significant challenge to society, but it can be particularly harmful to the individuals who become victims. This chapter engages in a comprehensive and topical analysis of the cybercrimes that target individuals. It also examines the motivation of criminals that perpetrate such attacks and the key human factors and psychological aspects that help to make cybercriminals successful. Key areas assessed include social engineering (e.g., phishing, romance scams, catfishing), online harassment (e.g., cyberbullying, trolling, revenge porn, hate crimes), identity-related crimes (e.g., identity theft, doxing), hacking (e.g., malware, cryptojacking, account hacking), and denial-of-service crimes. As a part of its contribution, the chapter introduces a summary taxonomy of cybercrimes against individuals and a case for why they will continue to occur if concerted interdisciplinary efforts are not pursued.\blfootnote{This is an article pre-print of the chapter: ``Cybercrime and You: How Criminals Attack and the Human Factors That They Seek to Exploit'' by Jason R.C. Nurse, due to appear in The Oxford Handbook of Cyberpsychology (2018/19), Edited by Alison Attrill-Smith, Chris Fullwood, Melanie Keep, and Daria J. Kuss. https://dx.doi.org/10.1093/oxfordhb/9780198812746.013.35.}

\keywords {cybercrime, cyber security, human psychology, cognitive science, social engineering, online harassment, hacking, malware, human factors} 
\end{abstract} 

\section{Introduction}
\label{sec:introduction}
% no \IEEEPARstart
\subsection{The Internet and Its Significance to us as Individuals}
Technology drives modern day society. It has influenced everything from governments and market economies, to global trade, travel, and communications. Digital technologies have further revolutionized our world, and since the advent of the Internet and the World Wide Web, society has become more efficient and advanced \cite{graham2014society} . There are many benefits of the online world and to such large scales of connectivity. For individual Internet users, instantaneous communication translates into a platform for online purchases (on sites such as Amazon and eBay), online banking and financial management, interaction with friends and family members using messaging apps (e.g., WhatsApp and LINE), and the sharing of information (personal, opinion, or fact) on websites, blogs, and wikis. As the world has progressed technologically, these and many other services (such as Netflix, Uber, and Google services) have been made available to individuals with the aim of streamlining every aspect of our lives.

In a 2017 study of 30 economies including the United Kingdom (UK), United States of America (US), and Australia, it was the citizens of the Philippines that spent the most time online---at eight hours fifty-nine minutes, on average, per day---across PC and mobile devices \cite{WeAreSocial2017}. Brazil was second with eight hours fifty-five minutes, followed by Thailand at eight hours forty-nine minutes online. Developed countries such as the US, UK, and Australia posted usage values of between six hours twenty-one minutes and five hours eighteen minutes. This highlights a substantial usage gap compared to some developing states. A key driver of this increased Internet usage is social media, and particularly individuals’ use of platforms such as Facebook, Facebook Messenger, WhatsApp, YouTube, and instant messaging service QQ \cite{WeAreSocial2017}. Evidence supporting this reality has also been found in other studies, where social networks are more frequently used by Internet users in the emerging world (Poushter, 2016); this type of use is key to understanding the impact of social media in online crime, as will be outlined further later in this chapter.

\subsection{The Prevalence of Cybercrime}
To critically reflect on today’s world, while the Internet has various positive uses, it is increasingly being used as a tool to facilitate possibly the most significant challenge facing individuals’ use of the Internet: cybercrime. Cybercrime has been defined in several ways but can essentially be regarded as any crime (traditional or new) that can be conducted or enabled through, or using, digital technologies. Such technologies include personal computers (PCs), laptops, mobile phones, and smart devices (e.g., Internet-connected cameras, voice assistants), but the scope is quickly expanding to encompass smart systems and infrastructures (e.g., homes, offices, and buildings driven by the Internet of Things, IoT).

The importance of cybercrime can be seen in its ever-rising prevalence. In the UK, for example, a key finding of an early Crime Survey of England and Wales by the Office for National Statistics (ONS) was that there were 3.8 million reported instances of cybercrime in the twelve months to June 2016 \cite{Scott2016}. This is generally noteworthy, but even more so, given that the total number of crimes recorded in the other components of the survey (e.g., burglary, theft, violent crimes, but excluding fraud) tallied 6.5 million. The number of cybercrimes, therefore, amounts to more than half of the total crimes. Similar trends can also be found in the 2018 ONS report, with cybercrime and fraud accounting for almost half of crimes \cite{techUK2018}. This reality becomes more concerning given that these statistics are only based on the reported crimes, and moreover, that such cybercrimes are almost certainly set to increase in the future. Studies from the US also further evidence the extent of cybercrime and identity theft. Research from the 2018 Identity Fraud Study found that \$16.8 billion was stolen from 16.7 million US consumers in 2017, which represents an 8\% increase in the number of victims from a year earlier \cite{Weber2018}.

\subsection{Types of Cybercrime}
At its core, there are arguably three types of cybercrime: crimes in the device, crimes using the device, and crimes against the device \cite{wall2007policing}. Crimes in the device relates to situations in which the content on the device may be illegal or otherwise prohibited. Examples include trading and distribution of content that promotes hate crimes or incites violence. The next category, crimes using the device, encompasses crimes where digital systems are used to engage and often, to deceive, victims. An example of this is a criminal pretending to be a legitimate person (or entity) and tricking an individual into releasing their personal details (e.g., account credentials) or transferring funds to other accounts. Wall’s final category, crimes against the device, pertains to incidents that compromise the device or system in some way. These crimes directly target the fundamental principles of cybersecurity, i.e., the confidentiality, integrity, and availability (regularly referred to as the CIA triad) of systems and data. This typology provides some general insight into the many crimes prevalent online today.

This chapter aims to build on the introduction to cybercrime and security issues online and focus in detail on cybercrimes conducted against individuals. It focuses on many of the crimes being conducted today and offers a topical discourse on how criminals craft these attacks, their motivations, and the key human factors and psychological aspects that make cybercriminals successful. Areas covered include social engineering (e.g., phishing, romance scams, catfishing), online harassment (e.g., cyberbullying, trolling, revenge porn, and hate crimes), identity-related crimes (e.g., identity theft and doxxing), hacking (e.g., malware and account hacking), and denial-of-service (DoS) crimes.

\section{Cybercrimes against Individuals: A Focus on the Core Crimes}
\label{sec:background}
The cybercrime landscape is enormous, and so are the varieties of ways in which cybercriminals can seek to attack individuals. This section introduces a taxonomy summarizing the most significant types of online crimes against individuals. These types of cybercrime are defined based on a comprehensive and systematic review of online crimes, case studies, and articles in academic, industry, and government circles. This includes instances and cases of cybercrime across the world (e.g., \cite{bbc2016a,Sidek2017}), taxonomies of cybercrime and cyberattacks that have been developed in research (e.g., \cite{gordon2006definition,wall2007policing,wall2015internet}), industry reports on prevalent crimes (e.g., \cite{CheckPoint2017,PwC2016}), and governmental publications in the space (e.g., NCA \cite{NCA2017}).

\begin{figure*}[ht]
	\centering
	\includegraphics[width=\linewidth]{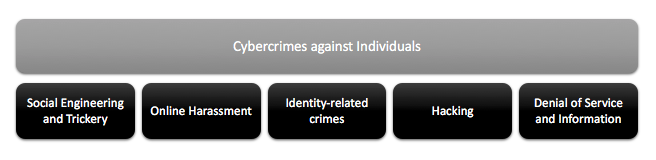}
	\caption{Main types of cybercrimes against individuals}
	\label{figone}
\end{figure*}

The intention is to connect the identified types of cybercrime to real-world situations, but also to maintain a flexible structure as new types of cybercrimes may well emerge. Moreover, the chapter is inclusive in its approach and defines types that are relatable and easily communicated—which has benefits for engagement, especially for those not involved in cybersecurity nor with a technical background or expertise. It is important to note here that many of the types identified here can be seen across prior works. For example, Wall’s work \cite{wall2015internet} examines crimes against the individual, crimes against the machine, and crimes in the machine, and Gordon and Ford \cite{gordon2006definition} use some of these types as exemplars of their Type 1 and Type 2 cybercrimes. This taxonomy’s value is therefore not in identifying new types of cybercrime, but instead in providing a new perspective on the topic which centers in on the types of cybercrime most prevalent today. The taxonomy is presented in Figure~\ref{figone}.

The first type of cybercrime is Social Engineering and Trickery, which involves applying deceitful methods to coerce individuals into behaving certain ways or performing some task. Next, Online Harassment is similar to its offline counterpart and describes instances where persons online are annoyed/abused and tormented by others. Identity-related crimes are those in which an individual's identity is stolen or misused by others for a nefarious or illegitimate purpose (e.g., fraud). Hacking, one of the most well publicized cybercrimes both in the news and the entertainment industry (e.g., Mr. Robot, Live Free or Die Hard, The Matrix, Swordfish), is the action of compromising computing systems. While traditionally not regarded as a significant personal crime, Denial of Service is one of the most used by online criminals, and its popularity is attributed to its simplicity---i.e., it primarily involves blocking legitimate access to information, files, websites, or services---and effectiveness. Finally, (Denial of) Information accommodates the new trend of ransomware which in similar in that it denies individuals access to their own information. The next sections analyze the taxonomy and each of its types of crimes in detail.

\section{Social Engineering and Online Trickery}
Trickery, deceit, and scams are examples of some of the oldest means used by adversaries to achieve their goals. In Greek Mythology, their army used deceit in the form of a Trojan horse; presented to the Trojans as a gift (or more specifically, an offering to Athena, goddess of war), it was instead a means for the Greek army to enter and destroy the city of Troy. Additionally, in The Art of War, fifth-century BCE Chinese military strategist Sun Tzu declares, ``Hence, when able to attack, we must seem unable; when using our forces, we must seem inactive; when we are near, we must make the enemy believe we are far away; when far away, we must make him believe we are near'' \cite{tzu2009}. According to this well-known text on war, the intention is to deceive and, ideally, to misdirect, while discretely progressing towards and obtaining the goal---in Tzu's case, winning against the enemy in battle.

Cybercriminals, potentially informed by history itself, have been applying such techniques for decades to Social Engineering, a specific class of cybercrime that uses deception or trickery to manipulate individuals into performing some unauthorized or illegitimate task. It seeks to exploit human psychology and is possibly the most effective means of conducting a crime against an individual.

In one example, a social engineer breaks into an individual’s cell-phone provider account in under two minutes \footnote{https://www.youtube.com/watch?v=lc7scxvKQOo}. This was achieved by phoning the cell-phone provider’s help desk, pretending to be the customer’s wife (impersonation is typically a core component of this crime), and using an audio recording of a crying baby (under the guise of it being her baby) to elicit sympathy from the help desk employee. Here, the social engineer used some basic information (i.e., knowing the customer's name), sympathy, and the fact that a help desk is primarily supposed to provide assistance, to manipulate the help desk to grant her unauthorized access to a client account. There are numerous other similar types of attacks, and entire books (e.g., \cite{hadnagy2010social,mann2008hacking,mann2013hacking}) and training courses on the topic (e.g., at the well-known hacking conference, BlackHat). 

\subsection{Phishing and Its Variants}
Phishing is a specific type of social engineering crime that occurs using electronic communications, such as an email or a website. In it, criminals send an email, or create a website, that appears to be from a legitimate entity with the intention of conning individuals into divulging some sensitive information or performing a particular action. Today there are many different variants of phishing, including spear-phishing, vishing, smishing (or SmShing), and whaling.

Spear-phishing is a targeted phishing attack on an individual that has been customized based on other key and pertinent information, such as their date of birth, current bank, Internet service provider, or email address. This additional information is used to enhance the appearance of legitimacy and thereby increase the effectiveness of the con. Spear-phishing is held to be the reason for several well-known crimes including ``Celebgate'', where private photographs of actresses Jennifer Lawrence, Kate Upton, and Scarlett Johansson were stolen and later exposed online. The terms vishing and smishing represent phishing attacks that occur over the phone (i.e., voice), and via text messages (especially SMS, but including WhatsApp, etc.) respectively. These often overlap with traditional phone scams but may also be used in combination with email phishing attempts. Whaling is very similar to spear-phishing but targets high-profile individuals (the notion being that a whale is a ``big phish'') such as company executives, with the goal of a higher payoff for criminals if the attack is successful.

The success of phishing attacks over the last decade has been phenomenal. To take the UK as an example, the City of London Police's National Fraud Intelligence Bureau (NFIB) and the Get Safe Online security awareness campaign estimated that in 2015 alone, phishing scams cost victims \pounds174 million. Moreover, Symantec \cite{Symantec2017} estimates that spear-phishing emails as a category in themselves have drained \$3 billion from businesses over the last three years. These estimates are likely to increase, as are the various ways in which criminals have targeted individuals.

In one phishing scam, criminals monitored a lady in the process of purchasing a home, and after disguising themselves as her solicitor they requested that she transfer \pounds50,000 into their account \cite{itv2015}. This can be considered as a spear-phishing attack given the amount of information the criminals had on her and her activities, and how they used that information to achieve their goal (similar to the process of reconnaissance). There have also been emails sent to university students where criminals have posed as employees of the university’s finance department. They pretend to offer educational grants that can only be redeemed after students provide personal and banking details \cite{bbc2016a}. While emails are prominent tools, fake websites also are a popular avenue for phishing crimes. A 2017 study discovered hundreds of fake websites posing as banks, including HSBC, Standard Chartered, Barclays, and Natwest, that targeted the public \cite{McGoogan2017}. These websites looked identical to official sites and used similar domain names, such as $hsbc-direct.com$, $barclaya.net$, and $lloydstsbs.com$ (note the additional letter or slight re-organization of bank name in these addresses).

A key observation about these attacks and those above is that criminals have sought to exploit many human psychological traits. These include a willingness to trust others and to be kind, the impact of anxiety and stress on decision making, personal needs and wants, and in some regards, the naivety in decision making. In the home purchase case, criminals firstly targeted the stressful process of purchasing a home, and then secondly, waited for a specific moment in time where they could impersonate the solicitor to request transfer of funds. While not privy to the email sent, the tone of the email must have emphasized the importance of transferring the funds immediately to secure the purchase. Fear of losing the prospective property, the overall anxiety of house buying, and trust in the (supposed) solicitor are undoubtedly factors that would have led to the transfer of funds. Mann \cite{mann2008hacking} mentions similar tricks as core to social engineering, and Iuga, Nurse, and Erola \cite{iuga2016baiting} mention these tricks as increasing the susceptibility of individuals to phishing attacks.

In the case of the university students, criminals targeted a prime need of students during their time at university, i.e., financial support to fund their degrees and themselves. By using university logos and other information, they were able to pose as a legitimate entity and thereby not arouse the suspicion of students. This impersonation also occurs within the fake website example. Criminals prey on naïve decision-making abilities, or more specifically, the heuristics (or quick “rules of thumb”) that individuals apply to make decisions. Here, they are presenting emails and sites as we expect they should appear, thus deceiving us into accepting them and acting without detailed consideration. This process has previously been described via the psychological heuristic of representativeness by psychologists Tversky and Kahneman during the 1970s. The heuristic posits that humans often make decisions based on how representative an event is grounded on the evidence, rather than what may be probabilistically true \cite{kahneman1973psychology}. Therefore, because the website or email appears to possess all of the key evidence (a logo, familiar names, etc.), its legitimacy is more likely to be accepted. This is only one example of the ways in which psychology overlaps with cybersecurity; many others can be found in Nurse, Creese, Goldsmith, and Lamberts \cite{nurse2011trustworthy}.

\subsection{Online Scams---Tech Support, Romance, and Catfishing}
In addition to phishing, online scams are also worth mentioning. Scams also involve trickery and deceit and typically have financial gain as the prime motive. One prominent example of the now common series of ``tech support'' scams is that of a global con uncovered in 2017. There, criminals purchased pop-up browser advertisements which appeared on victim’s computer screens and locked their browsers \cite{DoJ2017}. These pop-ups inaccurately informed individuals that their computers were compromised and that they should call the “tech support” company for assistance. Reports indicate that over 40,000 people across the globe were victimized and defrauded out of more than \$25 million USD \cite{DoJ2017}. These criminals were using a series of fear tactics to deceive individuals, many of whom were elderly and potentially more vulnerable.

Romance scams are also rampant on the Internet via online dating websites. Here, criminals seek to engage in faked and extensive relationships, again, usually for financial gain. Their technique involves preying on vulnerable individuals seeking romance and love and exploiting them under the guise of a relationship. Research has studied these scams from a variety of perspectives, including understanding their prevalence (e.g., \cite{whitty2012online} and their impact on victims (e.g., \cite{whitty2016online}. A noteworthy finding for our work on cybercrimes and individuals is that while financial losses may be incurred by victims, it is often the loss of the relationship that was more upsetting and psychologically traumatic. Catfishing is another variant of the common romance scam where fake, online identities and potentially, even social groupings are created to lure individuals into romantic relationships. Similar to traditional scams, the goal may be for financial gain, but notoriety may also be considered as a motive, e.g., American football player Manti Te'o \cite{schulman2014real}. Te’o was famously tricked into believing that he was in a relationship with Stanford University student Lennay Kekua, who, in reality, did not exist: Te'o was the victim of a year-long girlfriend hoax.

It is also important to consider the reasons behind why people continuously fall for online scams in the face of the large amounts of publicity to educate and warn individuals. Although fear, trickery, and the targeting of vulnerable individuals all play large parts, other research has extended consideration of these issues. Button, Nicholls, Kerr, and Owen \cite{button2014online} have also identified core motivational factors that include the diversity of scams and frauds (i.e., criminals may find areas where individuals may be less wary of being defrauded), small amounts of money sought by criminals (if small amounts of money are lost, this may worry individuals less), authority and legitimacy displayed by scammers (this touches on the previous point of trickery and impersonation), as well as visceral appeals (i.e., criminals devising scams that appeal to human needs/feelings such as finance, love, sex, and sorrow). These cut across the various scams covered here and provide some insight into the diverse ways criminals use trickery and social engineering to achieve their nefarious goals, and thus why scams continue to be successful.

\section{The Challenge of Online Harassment}
Online harassment can broadly be regarded as the targeting of individuals with negative terms or actions. Emphasizing the significance of this crime, a 2016 Data \& Society Research Institute study found that that 47\% of U.S. Internet users have personally experienced online harassment or abuse, and 72\% of these users have seen someone harassing someone else online \cite{Lenhar2016}. In terms of types of individuals that have been targeted, the research found that men and women are equally likely to face harassment online, but the latter have experienced a wider diversity of abuse. The individuals that are more likely to experience or witness abuse online include young users, black users, or those that identify as lesbian, gay, and bisexual (LGB). These findings broadly demonstrate an upwards progression from 2014 research by Marie Duggan at the Pew Research Center that also focused specifically on understanding online harassment \cite{PEW2014}.

In the UK, statistics collated by the National Society for the Prevention of Cruelty to Children (NSPCC) indicate a similarly worrying situation, especially considering children and online abuse. They note that one in three children have been victims of bullying online and almost one in four young people have come across racist or hate messages online \cite{NSPCCnd}. According to the NSPCC, such harassment has led to over 11,000 counselling sessions with young people who talked to ChildLine (a U.K. help and advice hotline) about online issues between 2015 and 2016.

\subsection{Cyberbullying}
Cyberbullying is one of the various types of online harassment, and one of many that are online manifestations of offline malevolent actions. It affects children, teenagers, and adults alike. It, like bullying, essentially involves repeated aggression (direct or indirect) levied by a group or individual against a victim that is (often) unable to easily defend him/herself. This aggression however, now occurs through modern technological devices such as the Internet or smartphones \cite{slonje2008cyberbullying}. There are countless examples of this crime to be found in the media and, tragically, a number of resulting instances of suicide among youth (e.g., \cite{bbc2016b,Turner2017}). A 2016 BBC report referred to one victim and noted that ``His confidence and self-esteem had been eroded over a long period of time by the bullying behavior he experienced in secondary education. People who had never even met [ … ] were abusing him over social media and he found that he was unable to make and keep friends'' \cite{bbc2016b}. This example captures the essence of cyberbullying, and also highlights the use of current platforms such as social media as one of its core conduits.

Research also contributes significantly to understanding the problem of cyberbullying. For instance, Whittaker and Kowalski \cite{whitty2016online} found that texting and social media are two of the most common venues for cyberbullying in college-age students. More interesting however, is the finding that there may be an overlap in roles between ``bully'' and ``victim'' and that despite the significant emotional impact of cyberbullying, many victims do not seek support \cite{price2010cyberbullying}. These factors are important because they suggest a continuation of cyberbullying due to related behavior, and the lack of treatment (which potentially leads to exacerbation). A key factor to point out here, as compared to social engineering, is that perpetrators are usually not conventional criminals. Instead, they tend to be individuals who do not recognize the full extent of the psychologically detrimental impact of their actions. This is especially the case with young people, where there may be a lack of awareness of others' feelings compounded by the inherent immaturity present in this age group. Cyberbullying is, however, also prevalent in adults (e.g., in social media and the workplace \cite{privitera2009cyberbullying}) even though the expectation exists for adults to be better informed and more cognitively aware of their actions than are young people.	

\subsection{Internet Trolling and Cyberstalking}
Internet trolling and cyberstalking are two other forms of online harassment that both share a few similarities with cyberbullying. Trolling is the action of posting inflammatory messages deliberately with the intention of being disruptive, starting arguments, and upsetting individuals. Bishop \cite{bishop2014dealing} identifies twelve types of ``trollers'' split into four groups: Haters (inflame situations for no benefit to others); Lolcows (provoke others to gain attention); Bzzzters (chat regardless of accuracy or value of contribution); and Eyeballs (wait for the opportune moment to post provocative messages). The motives for such actions have been empirically studied and relate to boredom, attention-seeking and revenge, fun and entertainment, and damage to the community and other people \cite{shachaf2010beyond}. This research provides useful insight into the types of actions that are core to trolling, and the motives of individuals who engage in it.

Real-world examples of trolls can be found in media reports and include people who have used online means such as social media to falsely brand others as pedophiles and witches, and also threatened to harm them \cite{Nunn2014,Telegraph2015p}. As a result of such online malfeasance, the UK is an example of a country that now has stringent laws regarding this behavior (notably the Malicious Communications Act) and has already sentenced several trolls to jail.

Cyberstalking is the use of electronic means (e.g., Internet, email) by criminals to repeatedly harass, threaten, prey on, or otherwise track an individual. Factors that tend to differentiate cyberstalking from other forms of online harassment include prolonged monitoring (or ``keeping tabs'') of victims and making victims feel afraid and unsafe. A more interesting distinction to consider, nonetheless, is what separates cyberstalking from offline stalking—which could assist in the understanding of its prevalence. Goodno \cite{goodno2007cyberstalking} defines five peculiarities exclusive to cyberstalking: cyberstalkers use electronic means to instantly harass victims and have opportunities for wide dissemination; they can be physically/geographically far away from their victims; criminals operate under a cloak of (perceived) anonymity online; they can easily impersonate their victims to aggravate situations; and finally, these cybercriminals often encourage third parties in their harassment. These differences are so significant that they have led to cyberstalking overtaking offline physical harassment in the UK as a crime \cite{McVeigh2011} .

While cyberstalking does affect a cross-section of society, research has shown that some groups and types of individuals are more likely to be targets. In one study for instance, LGB Internet users were found to be almost four times as likely to report experiencing continuous contact which made them feel unsafe \cite{Lenhar2016}. Women are also often targeted, e.g., for one female author, it had a serious impact on her personal and professional life \cite{gough2016}, and is one of many examples that illustrate how social media, in particular, can be used to support stalking. Here, the stalker continuously monitored the individual, tracked her movements, gathered personal data (e.g., her address), and contacted her son's school and newly met friends with malicious messages, e.g., from the stalker to a friend via Facebook—``One of the people around you is author [author’s name]. She seems like a nice person at first-but actually she is a toxic person under a silver tongued mask. [Author’s name] is a secretly sadistic narcissistic person who tries to get others to commit suicide. STAY AWAY FROM HER...She is a wolf in sheeps' clothing and has no conscience'' \cite{gough2016}. This example demonstrates one of the ways in which stalkers can use the Internet to abuse and control their victims, i.e., through targeting friends and family; this is in addition to the more direct forms of harassment (e.g., attempts at ongoing messages or persistent threats).

The challenge here is that the Internet and social media have become so embedded in the modern lifestyle that these technologies and individuals' tendency to overshare provides cyberstalkers and other criminals with copious amounts of personal information they need \cite{nurse2015exploring}. Additionally, Cavezza and McEwan \cite{cavezza2014cyberstalking} found that, compared to offline stalkers, cyberstalkers may be more likely to be ex-intimate partners. These results are interesting because they provide further insight into the types of people who perform such actions as well as those who are often impacted.

\subsection{Revenge Porn and Sextortion}	
Revenge porn and sextortion are two of the newest (in broad terms) forms of online harassment. Within the former, individuals, especially ex-partners, post sexual images of victims online without their permission. Criminals use these photo leaks to embarrass, humiliate, and demean victims. Sextortion is the gathering of sexual images or video (potentially via entrapment), and its use to blackmail individuals for further sexual footage or other favors. Reports indicate the significance of these crimes in cyberspace, with Facebook having to disable more than 14,000 accounts related to this form of crime in a single month alone \cite{hopkins2017}. Examples of these crimes can typically be found in two main scenarios.

The first scenario involves disgruntled ex-partners using private photos, likely shared during a previous sexual relationship, to humiliate their victims---this may occur especially if relationships did not end amicably. This has also become known as revenge porn, or more accurately, non-consensual sharing of private sexual images. Secondly, there are an increasing number of cybercriminal gangs using the guise of attractive young women to trick individuals into sexually explicit actions online (e.g., via webcams or Skype sessions). These actions are recorded and later used to blackmail victims---typically using threats of sharing photos with family and friends unless money is paid \cite{Sawer2016}.

Cybercriminals have also combined sextortion with phishing and hacked passwords to boost impact. The latest trend has been in emailing individuals claiming to have compromising video of them watching pornography, and recorded via their webcam; the email includes one of the individual’s passwords (attained most likely from a prior organizational data breach) to suggest legitimacy. Individuals are asked to pay a certain amount (e.g., via Bitcoin) or risk the video being sent to friends, family and coworkers. A poignant example, taken from the EFF, is as follows: ``Hi, victim. I write you because I put a malware on the web page with porn which you have visited. My virus grabbed all your personal info and turned on your camera which captured the process …. Just after that the soft saved your contact list. I will delete the compromising video and info if you pay me 999 USD in bitcoin. … I give you 30 hours after you open my message for making the transaction'' \cite{Quintin2018}.

Similar to the other crimes mentioned, revenge porn and sextortion can have devastating impacts on victims. In possibly one of the largest studies on the topic, Henry, Powell, and Flynn \cite{henry2017} found that 80\% of people who experienced sextortion reported heightened levels of psychological distress, such that it was also consistent with moderate to severe depression and/or anxiety disorder. Furthermore, victims often felt highly fearful for their safety after the ordeal. This response is well-justified as there have been other reports of serious threats (e.g., abuse and threats of rape) to victims of revenge porn \cite{Raven2014}, and other reports of suicide due to its prolonged effects \cite{bbc2017a}. It is worth mentioning that most research up until this point has focused on the legal and criminal aspects of revenge porn and combatting it. Simultaneously, there has been a surge in new laws (e.g., the U.K. Criminal Justice and Courts Act 2015, the Protecting Canadians from Online Crime Act) and subsequent prosecutions for criminals involved in these types of acts \cite{CPS2015}.

\subsection{Hate Crimes}	
Hate crimes (and hate speech) are another form of offline harassment that have made the transition to online. These are crimes that arise due to prejudice based on race, sexual orientation, gender, religion, ethnicity, or disability \cite{mcdevitt2002hate}. In many ways, these crimes overlap with those mentioned, and also extend them in terms of the threats levied. Jacks and Adler \cite{jacks2015proposed} build on earlier work (e.g., \cite{mcdevitt2002hate}) to examine the types of users that are engaged in online hate crimes (or with hate materials). They identify four main types: Browsers (viewers of hate material); Commentators (viewers and those who engage with and post comments); Activists (those who add overt hate material and seek to promote their views and engage with others); and Leaders (individuals who use the Internet to support, organize, and promote their extremist ideologies). As to be expected, Leaders are typically the smallest group, but as Jacks and Adler \cite{jacks2015proposed} note, they tend to be high repeat offenders.

Social media also plays a central role in hate speech and crimes, particularly those that occur after significant events. For instance, after the Woolwich attack on an off-duty soldier in London in May 2013, there were hundreds of hate messages posted on social media, especially Twitter, targeting Muslims \cite{awan2014islamophobia}. These perpetrators were using the platform of social media, and its wide reach, to openly attack people due to their faith. This issue of hate on social media has become so widespread that London's Met Police have set up an Online Hate Crime Hub unit to address it, and there have been demands for fines on Facebook, Twitter, and YouTube for failing to act swiftly against such content \cite{ArsTechnica2017}. It is arguably only via such concerted efforts that progress will be made in tackling the issue of online hate, but also that of online harassment more broadly.

\section{Identity-Related Cybercrime}
Identity theft and identity fraud are traditional crimes that have flourished due to online systems and the open nature of the Internet. While the theft of identities by criminals is enabled due the amount of information on individuals online, fraud becomes possible when that information is used for monetary gain (e.g., impersonating the individual to purchase an item). In the UK alone, there were just short of 173,000 incidents of identity fraud in 2016, which represents 53.3\% of all reported fraud, and more importantly, 88\% of this occurred online \cite{bbc2017b}. The U.S. market has also witnessed significant rises in identity-related fraud, with a 40\% increase in 2016 in ``card not present'' (i.e., mainly online) fraud \cite{jav2017} and in 2017, this type of fraud being 81\% more likely than point-of-sale fraud \cite{Weber2018}. These reports also act to highlight some of the main activities by cybercriminals engaging in identity theft and fraud, e.g., making online purchases, signing up for credit accounts (e.g., credit cards or loans), signing up to paid websites. Depending on the amount of data possessed by these criminals, there are even concerns that they could apply for passports in a victim’s name. Other examples of crimes such as unlawful identity delegation and exchange have also been documented in research \cite{koops2009typology}.

Identity theft works by criminals gathering information on individuals and using that as the basis through which to steal their identities. Today, there are two information-gathering techniques preferred by cybercriminals: the monitoring of individuals on social media as they post and interact online, and the gathering and use of personal data from previous online security breaches. The first of these techniques exploits a factor previously mentioned that pertains to phishing, i.e., the nature to overshare, but also the poor management of security and privacy online. A noteworthy study by fraud prevention organization Cifas found that Twitter, Facebook, and LinkedIn are now prime “hunting grounds” used by identity thieves \cite{Samee2016}; these networks contain an abundance of personal details, from birth dates and family member details to addresses, school histories, and job titles.

Previous research has considered this issue of oversharing and modeled how social media data could be used to place individuals at great risk, both online and offline \cite{creese2012data,nurse2015exploring}. There are also greater impacts on security and privacy as this data is combined with that from IoT devices such as fitness trackers and smart watches \cite{aktypi2017unwinding}. Most recently, people using Strava to track their exercise patterns inadvertently exposed details of military bases when posting their results to the app; such types of exposure can increase the risk to individuals, businesses, and governments \cite{hern2018,nurse2018strava}. In addition to focusing on these risks, other relevant psychological research has sought to understand why individuals tend to disclose more online. This has led to the identification of six factors which explain such behavior and create what has been deemed the ``online disinhibition effect'': dissociative anonymity (separation of online actions from offline identities); invisibility (opportunity to be physically invisible and unseen); asynchronicity (lack of immediate and real-time reactions); solipsistic introjection (or, merging of minds with other online individuals); dissociative imagination (impression of the online world as make believe and not connected to reality); and minimization of status and authority (based on the perspective that everyone online is equal) \cite{suler2004online}. These factors, including their interactions, are widely considered to impact online behavior, and thus may also potentially be linked to exposure to risks (such as identity theft and fraud).

The second information gathering technique used by cybercriminals is that of previous online data breaches. Over the last ten years, a significant number of companies have been victims of cyberattacks and subsequently have leaked customer data online. A few well-known enterprises include Yahoo!, Uber, Target, Sony, Anthem (health insurer), JP Morgan Chase, Ashley Madison, and eHarmony, and the data exposed spans biographic information, medical records, email addresses, family members, social security numbers, card details, and passwords. These customer details have often been available openly on public websites (e.g., pastebin.com), or for sale online. $Pastebin.com$ provides an interesting case study given that although it has positive uses, hackers have become increasingly attached to it to publicly share/expose sensitive details (in addition to the above, this includes compromised social media accounts, access credentials to companies, etc.) online. Likely reasons for this preference include the site's lack of requirement for users to register, its lack of proactive moderation of posts, and its ability to handle large text-based files.

The Dark Web is particularly relevant here as it is one of the most well-known places where identity data and banking details can be found and traded by cybercriminals. Because the Web exists on an encrypted network it can only be accessed by tools such as Tor (The Onion Router), and thus offers some level of anonymity. According to the Underground Hacker Marketplace report, credit cards can be purchased for as little as \$7 USD, identity packages (including social security number, driver’s license, and matching utility bill) for \$90 USD, and a dossier of credentials and data (dubbed a Fullz, and containing names, addresses, banking information, and physical counterfeit cards) for \$140-\$250 USD \cite{Dell2016}. Such cybercrime marketplaces and ecosystems place individuals at a continued risk of identity theft and fraud, especially considering that much of an individual’s most valued identity data (e.g., name, email, social security number, bank accounts) is not easily changed.

Although it is not as significant (at least from a monetary standpoint) as identity theft or fraud, the newer crime of online doxxing (or doxing) is worth a mention here. This attack involves inspecting and researching personal information (e.g., home addresses, emails and phone numbers, preferences) about an individual and then posting that information publicly online. The criminal’s intention is generally to infringe on the privacy of that person for malicious reasons such as harassment, or to conduct some form of vigilante justice for an actual or perceived wrong.

\section{Hacking: The Dark Art}
Hacking is one of the most traditional forms of cybercrime and involves activities that result in the compromise of computing systems and/or digital information. By compromise, this chapter refers specifically to the detrimental impact of these actions on the confidentiality and integrity of systems and data. As such, hacking can refer to corporate or personal data (e.g., a person’s photo album) being exposed, or accessed by, unintended parties; the unauthorized modification or deletion of that data (with or without the knowledge of the individual); or computer systems being disrupted from functioning as intended.

\subsection{Malware (Viruses, Worms, Trojans, Spyware, and Cryptojacking)}
There is a plethora of crimes that can be labeled as hacking. The most topical threat in this domain however, is arguably that of malware. Malicious software (or malware) describes applications developed and used by criminals to compromise the confidentiality or integrity of systems and information. The cost of managing malware alone for U.K. organizations in a 2016 study totaled \pounds7.5 billion \cite{Warwick2016}. This has been matched by an even more drastic increase in the amount of malware applications and variants deployed by criminals. For instance, in 2017, Symantec \cite{Symantec2017} reported a threefold increase in new malware families online, while in 2018 there was a 88\% increase in new malware variants (\cite{Symantec2018}. The most popular types of malware that impact individuals are viruses, worms, Trojan horses, and spyware.

Viruses are programs that replicate when executed and spread to other files and systems. They are known for attaching themselves to other programs. The Melissa virus is one of the most famous viruses in history. It was implemented as a Microsoft Word macro virus that once opened by an unwitting individual, automatically distributed itself via email to the first 50 people in that individual’s Outlook address book, with the message ``Here is that document you asked for...don't show anyone else ;-)''. As these emails were opened and the document was accessed, the virus would spread even further, infecting more computers, and generating thousands of unsolicited emails. A unique characteristic of Melissa (and many of the viruses it has since inspired) was that its success and the continued spread of the virus exploited human psychology. Specifically, it targeted individuals’ friendships, i.e., sending to contacts thereby hijacking existing trust relationships, and also used trickery by referencing a document that was supposedly requested and allegedly secretive.

Worms are similar to viruses but they are standalone and do not need to be attached to a file. The prime purpose of worms is to self-replicate especially to other computers on the network (e.g., a home, university, or public network). As a result of its purpose, worms tend to vastly consume system resources (e.g., a computer’s CPU and memory, and a network’s bandwidth) thus slowing down computers and network speeds. Examples of recorded computer worms include Blaster, which would also cause the user’s computer to shut down or restart repeatedly, ILOVEYOU, and the Daprosy worm.

Trojan horses, as the name suggests, are programs that appear legitimate but have another core purpose, which commonly is acting as a back door into computers or systems (most notably, Remote-Access Trojans (RATs)). These malware variants can allow cybercriminals to circumvent security mechanisms to gain unauthorized access into systems. This access may be used to steal files, monitor individuals, or to employ the computer as a proxy for a larger attack. For example, personal information and files (e.g., photo albums, information on finances, private diaries, saved passwords) may be accessed and leaked online, or criminals may remotely turn on web cameras to spy on and take photos of individuals (e.g., \cite{Engadget2016,Korolov2016}). The latter of which could lead to sextortion. Furthermore, computers could be used as a platform to launch cyberattacks against other systems. This is similar to the recent case of the DoS attack on DNS provider, Dyn, where IoT devices from within homes and organizations across the world aided in disrupting access to hundreds of popular websites \cite{Krebs2016}.

Another type of malware targeting individuals online is spyware, which, as the name suggests, spies on and collects information about users, which could span from gathering specific information (e.g., passwords, banking information, search habits, computer-usage information) to storing all of the individual’s behavior on the computer or system. The primary goal of spyware is to extract useful information about users that can then be used by the cybercriminal for a financial gain. There are numerous instances of such malware found on computers and smartphones (e.g., \cite{CheckPoint2017,Lecher2016}).

While many of the other malware types have been known for some time, a more recent entry in the malware domain is that of cryptojacking typically through coin mining malware. Cryptojacking is the process of using an individual’s computing device (PC, laptop, etc.) without their knowledge to `mine' cryptocurrencies such as Bitcoin. Mining is a computationally expensive problem, and therefore, cybercriminals have sought to use any resources they can find---including hijacking the processing power of unsuspecting user devices---and pool these together to form a remotely linked system for efficient mining. This hijacking typically works by the hacker secretly including mining scripts (pieces of programming code) within webpages or browser extensions which automatically execute when a user visits a website. In early 2018, several government websites in the UK, US, and Australia were compromised by cryptojacking malware \cite{Osborne2018}, which meant visitors to those sites unwittingly may have participated in mining. Numerous other companies, networks and online sites have also been compromised by this threat, including Tesla, GitHub, a Starbucks Wi-Fi network, and a series of pirate video streaming websites. More worryingly, the problem of cryptojacking is likely to become significantly worse in the future as current reports note that attacks in the UK alone have surged 1200\% \cite{Martin2018} and over the course of 2017, there was a 34000\% increase in coin mining attacks \cite{Symantec2018}---the motivation for attackers being new currency or simply, more money.

Having reflected on the several types of malware present, it is also worthwhile to consider the ways in which individuals’ technology become infected, and thus what makes such crimes/attacks truly successful. Focusing on viruses and worms first, these are unique as they self-replicate and automatically spread to other systems with little user contact. The computers and users that are initially infected are therefore the key to the prevalence of this computer attack. Trojans horses, spyware, and their variants (e.g., adware and scareware) offer a different challenge to cybercriminals as to how they disseminate their attacks. There is a range of techniques developed to threaten individuals.

Phishing (and spear-phishing) attacks are the most common vector through which criminals transmit malware \cite{Symantec2017}. These exploit the trust of humans through impersonation and social engineering. Another infection vector is the bundling of malware with legitimate software downloads; this regularly occurs with spyware and third-party browsers or applications such as peer-to-peer file sharing platforms like Kazaa \cite{moshchuk2006crawler}. Here, cybercriminals recognize the importance of certain applications and seek to exploit that by pairing installations. In many cases the pairing of additional software may not be known by users, although in some cases it may be and users may still choose to download it. From a psychological perspective, this may occur for multiple reasons. For instance, users may be focused only on their end goal (e.g., watching a film or listening to music) and ignore anything that distracts from that goal, or they may not want to pay for services and so prefer to watch a film online for free. There is also the reality that users often misunderstand the level of risk they are facing and overestimate the capability of protection measures such as anti-virus software \cite{nurse2011trustworthy}. This results in overly risky decisions, and ultimately may lead to the successfulness of a hack.

Watering hole attacks and drive-by downloads are also highly preferred techniques, and these demonstrate how simple it is to compromise individuals. These attacks only require individuals to visit an infected webpage or misclick in a browser window, and the malware will be downloaded automatically for later installation. Watering hole attacks are particularly interesting because they involve the cybercriminal monitoring the types of sites an individual or certain group tends to visit, and then compromising (one or more of) those sites to allow for the injection of malware (in essence, ``poisoning the watering hole''). They then wait until the intended targets visit those sites again and thus become infected. This exemplifies one of the many tailored attacks levied by cybercriminals to target individuals. It also demonstrates the research in which cybercriminals often engage and the extent to which they may be willing to monitor human behavior to increase successfulness of their crime. A crucial point worth noting here is that the sites targeted could be regular websites, and there is not necessarily an act, or fault, of the user that makes this attack possible other than visiting the site.

\subsection{Account and Password Hacking}
Beyond malware, the hacking of online accounts (e.g., Facebook, Gmail, Government portals, paid services) and user passwords is a significant challenge faced by individuals. This is due to a variety of techniques being applied by cybercriminals, many of which are now even automated. One popular approach to hacking an individual’s account is through the stealing of their username and password credentials. Criminals typically achieve this via shoulder surfing (i.e., looking over someone’s shoulder while they are entering their password), and cybercriminals also focus on installing malware on the victim’s computer that logs all keys typed (also known as a keylogger) or applying social engineering techniques.

A real-world example of such attacks was the case of a student who installed keyloggers on university computers to steal staff passwords, and then used their accounts to increase his test scores \cite{NakedSecurity2015}. Keyloggers are particularly dangerous as they can record all keystrokes, from passwords to credit card numbers. It is worth noting, however, that new approaches to stealing passwords are continuously being discovered, as evidenced with PINs deciphered through video recording and tracking the motion/tilt of smartphones \cite{mehrnezhad2017stealing,nurse2015smart}. The IoT could pose a real challenge here given the amount of personal information that may be leaked via the usage of smart devices---be they wearables (smart watches, fitness trackers), voice assistants (e.g., Amazon Alexa, Google Home, or Apple HomePod), or smart appliances (e.g., smart TVs, fridges, and ovens). Research has already demonstrated the somewhat irrational behavior of individuals when using the IoT, considering their beliefs regarding privacy versus their inaction to behave privately (i.e., the privacy paradox) \cite{williams2016perfect,williams2017privacy}.

Password guessing is another way in which cybercriminals can gain illegitimate access to individuals’ accounts. Informed guessing is the most successful technique and is where criminals use prior information to guess account credentials or infer details that would allow them to reset user accounts. Such information can be readily gathered from social media profiles (e.g., hobbies, pets, sports teams, mother’s maiden name, family member names, and dates of birth), which is why it is important for individuals to be wary of what they share online. Another avenue used by cybercriminals is that of previously breached passwords. Given the number of data breaches that have occurred over the last few years as discussed earlier and the tendency of individuals to reuse passwords across sites, criminals have the perfect platform to amass sensitive user data and existing credentials. Research has investigated this reality and demonstrated the various ways in which hackers can reuse and guess passwords with some degree of success using this prior knowledge \cite{das2014tangled}. Sites such as haveibeenpwned.com have since become popular as they allow users to check whether or not their account has been compromised in a breach.

Dictionary attacks, i.e., where words from the dictionary are used to form potential passwords, are also a common password hacking technique. Here, cybercriminals look to exploit poorly created passwords based on dictionary words. One unique aspect of these attacks is that they can be automated using hacking tools such as John the Ripper, Cain and Abel, and L0phtCrack. The availability of these tools, and the fact that they require little expertise yet combine several different password crackers into one packaged application, provides cybercriminals with a significant advantage. That is, that up-skilling and increasing the scale of attacks is much easier than before and thus, less of a barrier to conducting crime.

To exacerbate this issue, there are many common, weak passwords in use by individuals. A study of 10 million passwords sourced from data breaches that occurred in 2016 \cite{keepersecurity2017} highlighted several key points: firstly, the top five common passwords used by individuals were 123456, 123456789, qwerty, 12345678, and 111111; secondly, 17\% of users had the password ``123456.''; thirdly, the list of most frequently used passwords has demonstrated little change over the last few years; and finally, nearly half of the top 15 passwords are six characters or shorter. Fortune Magazine recently reported that many of these same issues occurred again in 2017 \cite{Korosec2017}. One inference that might be made from these findings is that users prefer to maintain simple and memorable passwords. This is hardly a surprise as security is often known to crumble when placed in conflict with usability \cite{nurse2011guidelines}, and after all, humans favor consistency and are known to be creatures of habit. For hackers, however, such weak and common passwords are ideal, and can be guessed extremely quickly, thus placing users at risk of account takeovers.

\section{Denial-of-Service (DoS) and Ransomware}
A DoS attack involves cybercriminals blocking individuals from accessing legitimate websites and services. This is normally achieved by bombarding the websites/services with an enormous number of fabricated requests (e.g., page visits), which causes legitimate requests to be dropped or the organization’s websites/services to crash under the load. This crime is somewhat unique as compared to the others above because it depicts another way that individuals may be impacted by cybercrime, i.e., via attacks on organizations and services that they use. Interestingly, there would be little obvious signs of this to a user other than the website being unavailable. Of course, the unavailability of a website does not necessarily mean a DoS attack has occurred; there are many other reasons that may be behind this, including human errors \cite{bbc2016c}.

On Christmas Eve of 2015 a DoS cyberattack inundated BBC services with a substantial number of web requests which eventually forced many offline \cite{Korolov2016}. While this attack was not unique (and, indeed there have been larger Distributed-DoS (DDoS) attacks, e.g., GitHub \cite{Kottler2018} or Dyn in 2016 \cite{Krebs2016}, there is one very worrying observation about it: the cybercriminals that claimed responsibility, a group named New World Hacking, stated that the attack was only a test and that they had not planned to take the BBC down for multiple hours. This demonstrates the power of cybercriminals today and suggests that, on occasion, they themselves are not fully aware of their capabilities. A compelling reason for this heightened and unknown capability might be the ease at which criminals can procure or rent hacking and botnet4 services on the Dark Web \cite{Dell2016}. Often, these services are rented without a proper understanding of their full impact.

In addition to DoS attacks, cybercriminals have also employed other forms of crime to block legitimate access requests by individuals. A popular trend today is using ransomware, which is a form of malware that encrypts individual's information and only allows subsequent access if ransom is paid (typically via the cryptocurrency, Bitcoin). Individuals might become infected by phishing attacks or using infected devices (e.g., pen-drives). According to Symantec \cite{Symantec2017} the growth of ransomware has been phenomenal, especially its use as a profit center for criminals. On average, they note that criminals demand \$1,077 USD per victim in each ransomware attack. There are many potential reasons for the growth in this crime, but arguably the most prominent is that criminals have fully recognized that an individual’s data, whether it be personal photos and videos, financial spreadsheets, or files, is their most valuable possession. As a result, these attacks are crafted to target that data.

The increasing prevalence of this crime is motivated by its high success rates. For example, 64\% of people in the US whose technology was infected were found to be willing to pay the necessary amount to regain access to their data \cite{Symantec2017}. Similarly, at an organizational level one infected hospital paid \$17,000 USD to have its files unencrypted \cite{Wong2017}. Psychologically, it is a simple decision of cost versus benefit for individuals and organizations: the cost of paying the ransom is significantly less than the benefit of having access to files, therefore the payment is made. For individuals, this might mean regaining access to precious videos of their child’s first steps or photos of a graduation or a selfie with a celebrity. For a hospital, access to the electronic health records database is required to be able to properly treat patients and thereby, to conduct business. Again, therefore, criminals have found a key weakness in these parties and are crafting crimes to carefully exploit them.

To further support their plight, cybercriminals are also making efforts to ensure that the paying of ransoms is as seamless and ``painless'' as possible. There have been anecdotes of cybercriminals providing ransom payment FAQs, helpdesks, and even offering discounts to individuals who cannot pay the full demands. This demonstrates a level of sophistication by criminals where crime is becoming an industry (see Nurse and Bada \cite{nurse2018bd}), capable of even offering ``customer services''. At the same time, there is an increasing amount of ransomware attacks, e.g., the WannaCry attack in 2017, which affected nearly 100 countries and critical services such as the U.K.’s National Health Service (NHS) \cite{Wong2017}. These attacks seem to increase due to the combination of reasons and raise a number of interesting questions for us as a society. For example, as these attacks continue to grow, will society simply accept them (and for instance, just pay the ransom)? Will the occasional (e.g., yearly) breach of our data simply be viewed as part of being online? And broadly, will we become desensitized (even further) to online risk? These present interesting avenues for future research in the field.

\section{Summarizing Key Human Factors, and Future Research}	
\label{sec:conclusion}
While the advantages that accompany Internet use and digital technologies are plentiful, there is an abundance of challenges and concerns facing the new, high-tech world. Cybercrime is one of the most prevalent and has the ability to impact people psychologically, financially, and even physically. This chapter reflected on many of the crimes that cybercriminals engage in today and the reasons why these are often quite successful, from social engineering and online harassment to hacking and ransomware attacks. A salient point is that cybercriminals are ready, willing, and have a strong history in exploiting many human psychological needs and weaknesses. Such facets include our innate desire to trust and help each other (e.g., in the case of the mother with the crying baby), the human need for love and affection (e.g., romance scams), the host of biases that affect decision-making on security \cite{nurse2011trustworthy}, and a perfect knowledge of what people consider most important, i.e., the willingness to pay for the return of something valuable (e.g., instances of ransomware). Table~\ref{label1} summarizes the main types of crimes and the respective human factors that may be exploited by cybercriminals to lead to their success.

%\bgroup
%\def\arraystretch{2}
\begin{table}[ht!]
	\caption{Types of cybercrimes and the respective human factors that are exploited by criminals}
	\label{label1}
	%{\normalsize{
	\begin{tabular}{|p{3cm}|p{11cm}|}
		\hline
		\textbf{Types of cybercriminal attacks}	& \textbf{Human factors that when exploited are likely to increase the crime’s success} \\ \hline
		Social Engineering and Trickery	& Individuals’ willingness to trust others, willingness to be kind or sympathetic, needs and wants (e.g., visceral appeals or desires for finances or help), suggested urgency or importance of a message (e.g., website or application prompt, email, or call) received (seeking to offset rational decision-making), signs of legitimacy or authority in a message or individual (e.g., branding identical to the official branding of individual or organization, with the aim of cultivating trust), fear as conveyed through a message or individual (meant to offset rational decisions), the targeting of situations that are high stress or where individuals are likely to be highly anxious (as in the case of the house purchase), convenience (where the easier decision may not be the most secure), and heuristics and biases (these overlap with many of the other factors).   \\ \hline
		Online Harassment	& Individuals’ tendency to overshare personal details online or trust an online identity too much to the point of exposing themselves (there is the potential for this contributing to specific targeting or harassment). There is also an indirect use of human factors by criminals, i.e., instead of relying on factors held by the victim, they also rely on the guise of their anonymity to launch their harassment (a perception that their real identities are hidden) and that they can encourage others to participate in the harassment. Forms of online harassment, such as sextortion, can also be combined with other crimes including phishing and hacking, to further panic victims and convince them to succumb to the criminal’s demands. \\ \hline
		Identity-related crimes	& Individuals' tendency to overshare personal identity details online, especially on forms of social media, including Facebook, Twitter, LinkedIn (this links human factors closely to the online disinhibition effect), and unfamiliarity with new forms of technology (new technologies such as the IoT may lead to further oversharing of identity data) which open individuals to risk. \\ \hline
		Hacking	& Individuals’ misunderstanding of how at risk they are (typically an underestimation), misunderstanding of the capability of security and privacy protection measures (often an overestimation), an individual’s wants and needs (for instance, bundling spyware with legitimate software), the emphasis on achieving goals potentially at the expense of security, tendency to overshare personal details online (which may lead to password guessing by hackers), selection of weak passwords because they are simple and memorable, and reuse of passwords across websites and applications (passwords which can often be gained from one of the hundreds of data breaches each year). \\ \hline
		Denial-of-Service (DoS) and Information	& Human factors in this context primarily relate to ransomware, and include: understanding the real value to an individual of their personal data (thus appreciating that the payment of a ransom is much less in value than that personal data, e.g., photos or financial information), and making the ransom payment process as seamless as possible (e.g., with FAQs, Helpdesks, and discounts).\\
		\hline
	\end{tabular}
	%}}
\end{table}
%\egroup

\clearpage

As the sophistication of cybercriminals has increased, so too must the approaches to prevent, detect, and deter their behaviors. Cyberpsychology research has made significant inroads to the analysis of this problem through the study of criminal behavior and the psychological and social impact on victims. The field of Cybersecurity features a range of new models, systems, and tools that aim to prevent and detect attacks against individuals---these utilize a variety of the latest techniques in machine learning and anomaly detection to boost accuracy and efficiency. Criminology is also a key area, and there are now several laws across the world seeking to deter online crimes and prosecute those who perpetrate them. However, if approaches towards preventing cybercrime are to be truly effective at protecting individuals, a more concerted, cross-disciplinary program is mandatory. It is only in this way that the insight from each field can be properly synthesized and combined to address the issue of online crime.

\bibliographystyle{splncs03}
\bibliography{bibliosocs}

\end{document}